\documentclass{article}
\usepackage{graphicx} 

\usepackage{authblk}
\usepackage{booktabs}
\usepackage{multicol}
\usepackage{multirow}
\usepackage{makecell}
\usepackage{xurl}
\usepackage{tabularray}
\UseTblrLibrary{booktabs}
\usepackage[top=0.8in,bottom=1in,inner=0.75in,outer=0.75in]{geometry}
\usepackage{hyperref}
\usepackage[export]{adjustbox}

\newcommand{\keywords}[1]{\noindent\textbf{Keywords---}#1}

\title{Bridging Cybersecurity Practice and Law: a Hands-on, Scenario-Based Curriculum Using the NICE Framework to Foster Skill Development}
\date{}

\author[1]{Colman McGuan\thanks{Corresponding author: \texttt{c.mcguan@vikes.csuohio.edu}}}
\author[1]{Aadithyan V. Raghavan}
\author[2]{Komala M. Mandapati}
\author[1]{Chansu Yu}
\author[3]{Brian E. Ray}
\author[4]{Debbie K. Jackson}
\author[2]{Sathish Kumar}

\affil[1]{\small Department of Electrical and Computer Engineering, Washkewicz College of Engineering, Cleveland State University}
\affil[2]{\small Department of Computer Science, Washkewicz College of Engineering, Cleveland State University}
\affil[3]{\small College of Law, Cleveland State University}
\affil[4]{\small Department of Instructional Excellence, Administration Center, Cleveland State University}

\begin{document}

\maketitle

\begin{abstract}
    In an increasingly interconnected world, cybersecurity professionals play a pivotal role in safeguarding organizations from cyber threats. To secure their cyberspace, organizations are forced to adopt a cybersecurity framework such as the NIST National Initiative for Cybersecurity Education Workforce Framework for Cybersecurity (NICE Framework). Although these frameworks are a good starting point for businesses and offer critical information to identify, prevent, and respond to cyber incidents, they can be difficult to navigate and implement, particularly for small-medium businesses (SMB). To help overcome this issue, this paper identifies the most frequent attack vectors to SMBs (Objective 1) and proposes a practical model of both technical and non-technical tasks, knowledge, skills, abilities (TKSA) from the NICE Framework for those attacks (Objective 2). The research develops a scenario-based curriculum. By immersing learners in realistic cyber threat scenarios, their practical understanding and preparedness in responding to cybersecurity incidents is enhanced (Objective 3). Finally, this work integrates practical experience and real-life skill development into the curriculum (Objective 4). SMBs can use the model as a guide to evaluate, equip their existing workforce, or assist in hiring new employees. In addition, educational institutions can use the model to develop scenario-based learning modules to adequately equip the emerging cybersecurity workforce for SMBs. Trainees will have the opportunity to practice both technical and legal issues in a simulated environment, thereby strengthening their ability to identify, mitigate, and respond to cyber threats effectively.

    \par\addvspace{\baselineskip}\keywords{Cybersecurity; SMB; NICE Framework; KSA; Curriculum}  
\end{abstract}

\section{Introduction}
\label{sec:introduction}
Cyberattacks are costly to businesses and are becoming more prevalent each year with 67\% of businesses reporting an increase in the number of cyber attacks experienced in the last 12 months as of 2024~\cite{ref:hiscox-cyber-readiness-2024}. Due to these cyberattacks, 43\% of the companies suffered a loss of customers and 47\% found difficulty attracting new ones, up from 21\% and 20\% in 2023~\cite{ref:hiscox-cyber-readiness-2024}. These attacks place immense financial pressure on businesses. In 2024, the cost of a data breach averaged \$4.88 million, a 10\% increase from 2023~\cite{ref:ibm-data-breach-report-2024}. Moreover, in 2024, 40\% of businesses identified their cyber resilience maturity as either \textit{basic} or \textit{ad hoc} in a self assessment---i.e., they lack formal processes and have limited cyber training and awareness~\cite{ref:hiscox-cyber-readiness-2024}. Businesses cite employees using their own devices as contributing to increased risk exposure~\cite{ref:hiscox-cyber-readiness-2024}, demonstrating the need for a robust and cross-functional cybersecurity skillset. This is especially concerning for small-medium businesses (SMBs) because of their limited resources and their lack of awareness~\cite{ref:dojkovski-fostering-information-security-culture, ref:paulsen-cybersecuring-small-businesses}. 

There are over 34 million SMBs in the US~\cite{ref:sba-2024-small-business-profile}, and cyberattacks are a serious threat to them. Cybersecurity awareness and education have played a significant role in improving employee awareness. This paper proposes a comprehensive model of knowledge and skillset that will assist SMBs in defending themselves. Our research was steered by three overarching questions.

\begin{enumerate}
    \item Where do SMBs stand with respect to cybersecurity? This question allows us to collect and synthesize existing data to identify the gap between the current SMB workforce and best practices.

    \item What are the most frequent and impactful attacks faced by SMBs? This allows us to center the scope of our research on providing the most robust skillset to SMBs. This is done to maximize the area of coverage with as small of a workforce as possible due to the limited nature of an SMB. Our research found that the most frequent attacks are phishing/social engineering, malware/ransomware, and web-based attacks (Section~\ref{subsec:identifying-the-most-frequent-attacks}).

    \item How can the workforce be equipped with the necessary knowledge and skills to apply the best practices? This helps us incorporate the NICE Framework~\cite{ref:nice-cybersecurity-workforce-framework} as a bridge between the workforce and the best practices by mapping the tasks, knowledge, skills, and abilities (TKSA) to the most frequent attacks. However, the NICE Framework lists a total of 634 knowledges, 377 skills, 1006 tasks, and 177 abilities, making it costly for SMB owners to investigate and implement. Our research yielded a total of 88 technical TKSA and 54 non-technical TKSA that are required to defend against the three previously mentioned attack types. This is 6.23\% of the total TKSAs present in the NICE Framework and is practical for SMBs to implement as it extracts the TKSA most relevant to defend against the attack vectors identified; by creating this subset of TKSA from the NICE Framework, SMBs are able to more efficiently target resources where necessary (Sections~\ref{subsec:identifying-best-practices-and-extracting-keywords} and~\ref{subsec:mapping-keywords-to-the-nice-framework}).

\end{enumerate}

After answering these questions, we develop a scenario-based training to enhance practical understanding and preparedness, which has been shown to be an effective learning model~\cite{ref:brilingaite-framework-for-competence-development-and-assessment-in-hybrid-cybersecurity-exercises}. The curriculum provides hands-on experience in simulated environments to improve students' ability to identify, mitigate, and respond to cyber threats as done in~\cite{ref:ghosh-assessing-competencies-using-scenario-based-learning-in-cybersecurity}. Guided by the NICE Framework, this curriculum prepares trainees, regardless of their academic background, to navigate the complex field of cybersecurity effectively.

The rest of the paper is organized as follows. Section~\ref{sec:preliminaries} provides brief preliminary information supplemental to understanding this work. Section~\ref{sec:related-work} discusses related works and touches briefly on the cybersecurity frameworks. Section~\ref{sec:proposed-work} discusses the proposed methodology behind our research including the identification of the three frequent attack types and determining the best practices to counter them and map them to the TKSA present in the NICE Framework. Section~\ref{subsec:scenario-based-curriculum} presents the design of the scenario-based learning model while Section~\ref{sec:results-and-analysis} analyzes students' performance and feedback. Lastly, Section~\ref{sec:conclusion} gives the closing remarks.

\section{Preliminaries}
This section provides supplementary information important for understanding this work.
\label{sec:preliminaries}

\subsection{A Note on KSA and TKSA}
\label{subsec:a-note-on-ksa-and-tksa}
KSA refers to knowledge, skill, and ability, and is the colloquial term used in our research field. However, the NICE Framework introduces tasks, creating the acronym TKSA. In this work, we tend to use KSA when speaking generally about the field of work due to its long-standing prevalence and well-understood meaning in the field. Conversely, when describing the NICE Framework keyword mapping, we tend to use TKSA as not to omit tasks when discussing our work.

\subsection{A Note on the NICE Framework}
This work is the culmination of several years of planning, research, and execution. As such, the most recent version of the NICE Framework at the onset of this work was NIST SP 800-181~\cite{ref:nist-nice-the-national-cybersecurity-workforce-framework}. Since then, NIST SP 800-181 Rev. 1~\cite{ref:nice-cybersecurity-workforce-framework-rev1} has been released. This section serves as an acknowledgment of the NICE Framework revision. In the future, we plan to compare the original version with the latest and alter our approach accordingly.

\subsection{Ethics Statement}
\label{subsec:ethics-statement}
The results of this work use human participation for data analysis. Their names are kept anonymous to protect their personal information.

\section{Related Work}
\label{sec:related-work}

There exist several cybersecurity frameworks; one such framework is the International Organization for Standardization (ISO) 27001, a well-known standard for information security management systems (ISMS)~\cite{ref:iso-information-security-cybersecurity-and-privacy-protection}. The NIST Framework for Improving Critical Infrastructure Cybersecurity consists of three main components: the \textit{Framework Core}, \textit{Implementation Tiers}, and \textit{Profiles}~\cite{ref:nist-framework-for-improving-critical-infrastructure-cybersecurity}. Factor Analysis of Information Risk (FAIR) Cyber Risk Framework~\cite{ref:freund-measuring-and-managing-information-risk}, is a model that helps businesses analyze, measure, and understand the risks posed by cyber incidents mainly in a two-way classification: loss event frequency and loss magnitude. Nonetheless, many businesses are still lacking in their understanding of these standards, particularly SMBs~\cite{ref:dojkovski-fostering-information-security-culture, ref:paulsen-cybersecuring-small-businesses, ref:chidukwani-a-survey-on-the-cyber-security-of-small-to-medium-businesses, ref:osborn-risk-and-the-small-scale-cyber-security-decision-making-dialogue}. According to the Organization for Economic Co-operation and Development (OECD), a medium-sized enterprise/business consists of an employee count greater than 50 but below 200-500, with the upper limit varying from country to country, while a small-sized enterprise/business consists of an employee count of less than 50~\cite{ref:oecd-glossary-of-statistical-terms}.

To address the gap in the workforce, we propose a TKSA model from the comprehensive NICE Framework published by the National Institute of Standards and Technology (NIST)~\cite{ref:petersen-workforce-framework-for-cybersecurity}. The NICE framework defines a set of building blocks named tasks, knowledge, skills, and abilities. These four blocks form the foundation for building different competencies, work roles, and teams. Businesses and individuals can use the framework to assess or train themselves---it acts as a bridge between educators, employees, and businesses.

The NICE Framework and the TKSA model have been the foundation of several cybersecurity research works. Kim et al. proposed identifying the commonality and differences among three different sectors---the government, academia, and private---with respect to TKSA~\cite{ref:kim-an-explorator-analysis-on-cybersecurity-ecosystem-utilizing-the-nice-framework}. Their research was conducted by performing an ontological qualitative analysis using archival data, which  is a limitation of their research as their findings might not reflect the current market. Nevertheless, their research provides excellent insight into how TKSA can be related to roles in different sectors. While this research was helpful, it only points out the relationship between the three different sectors and how interconnected they are. A major takeaway from their work is that it highlights the versatility and the unique application of the NICE Framework.

Bada et al. performed a case study on developing a cybersecurity awareness program for SMBs~\cite{ref:bada-developing-cybersecurity-education-and-awareness-programmes-for-small-and-medium-sized-enterprises}. They first performed a literature review based on certain keywords and studied the best practices in securing a business’ cyberspace. Their final program was heavily based on the existing London Digital Security Center (LDSC) program which consists of five primary areas and changes to the program were performed as per their findings and recommendations. While this is a step in the right direction with spreading awareness about cybersecurity in SMBs and businesses in general, their research did not focus on specific attacks or utilize the NICE Framework, which can prove to be an excellent bridge between the workforce and educational institutes.

Tobey et al. studied how applying competency-based learning can help with creating an industry-ready cybersecurity workforce~\cite{ref:tobey-applying-competency-based-learning-methodologies}. Their work discusses the difference between the outcome-based approach, which is the current approach followed by most fields of study, and the competency-based approach. The NICE Framework gives us the ability to create our own competencies based on the different TKSA~\cite{ref:wetzel-nice-framework-competencies-assessing-learners-for-cybersecurity-work}. A competency statement is made up of a combination of different TKSA; it is flexible and can be changed as per the needs. Tobey et al. go on to suggest that the NICE Framework is a good starting point for educational institutions to start basing their courses.

\section{Proposed Work}
\label{sec:proposed-work}
\begin{figure}
    \centering
    \includegraphics[width=0.5\textwidth]{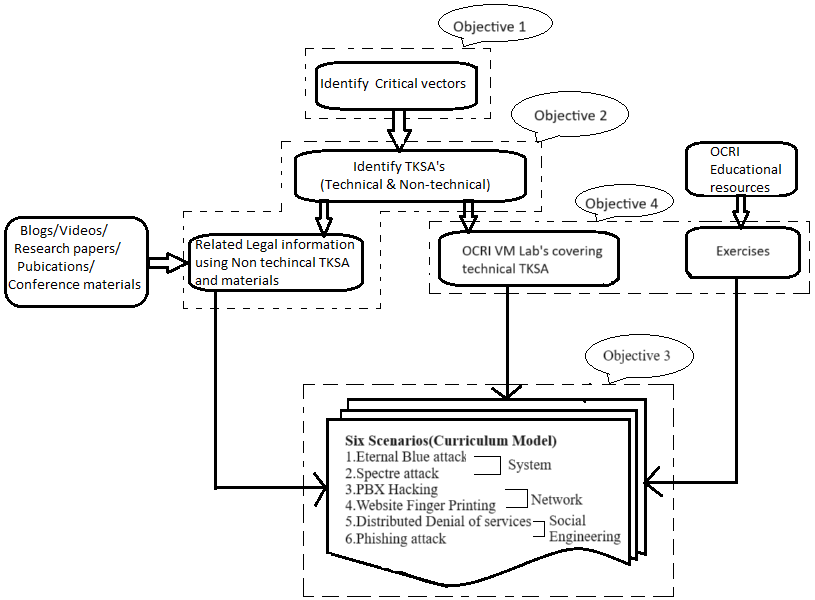}
    \caption{Flow Chart of the Curriculum Development.}
    \label{fig:curriculum-development-flow-chart}
\end{figure}

Our goal is to leverage technical and non-technical KSAs identified in the NICE Framework and map them to the most common cybersecurity attacks faced by SMBs. In doing so, we can design a hands-on and scenario-based curriculum to equip the emerging workforce with the legal and cybersecurity skills required to protect SMBs. Figure~\ref{fig:curriculum-development-flow-chart} gives a high-level overview of our approach.

This section is broken into four subsections. Section~\ref{subsec:identifying-the-most-frequent-attacks} discusses the attack identification process including the documents and reports used, and the justification behind selecting the attacks discussed in this paper. Section~\ref{subsec:identifying-best-practices-and-extracting-keywords} discusses the process of identifying the best practices to combat the attacks identified in Section~\ref{subsec:identifying-the-most-frequent-attacks}. This section also delves into the keyword extraction program used in this paper and explains the de-duplication of similar keywords, specifically the use of Levenshtein similarity. Section~\ref{subsec:mapping-keywords-to-the-nice-framework} shows the mapping of keywords to the NICE Framework and Section~\ref{subsec:integration-of-KSAs-virtual-machine-labs-and-NICE-framework} gives the basis of our scenario-based curriculum design.

\subsection{Identifying the Most Frequent Attacks}
\label{subsec:identifying-the-most-frequent-attacks}
To identify the most common attacks, we referred to reports published by Verizon, Hiscox Group, Ponemon Institute, and government agencies such as the Cybersecurity and Infrastructure Security Agency (CISA) and the European Union Agency for Cybersecurity (ENISA). They publish the latest trends of various cyberattacks, cost of mitigation, and other valuable information to help organizations prioritize their resources to defend themselves. The data they collect is usually based on surveys that multiple organizations participate in. Table~\ref{tab:list-of-documents-to-identify-attack-vectors} shows the list of the reports.

\begin{table}[htbp]
    \centering
    \begin{tblr}{vspan=even, colspec={Q[l]X[l]}, rowsep=2pt, width=\linewidth, rows={m}}
        \toprule
        Publisher & \SetCell[]{c}Document \\
        \toprule
        Verizon & Data Breach Investigation Report (DBIR) 2022~\cite{ref:verizon-2022-data-breach-investigation} \\
        \midrule
        Ponemon Institute & Ponemon Institute 2019 Global State of Cybersecurity in Small to Medium-sized Businesses~\cite{ref:ponemon-2019-global-state-of-cybersecurity} \\
        \midrule
        CISA & CISA Insights~\cite{ref:cisa-insights-ransomware-outbreak} \\
        \midrule
        Hiscox Group & Hiscox Cyber Readiness Report 2022~\cite{ref:hiscox-cyber-readiness-2022} \\
        \midrule
        \SetCell[r=2]{l}ENISA & Cybersecurity for SMEs---Challenges and Recommendations~\cite{ref:enisa-cybersecurity-for-smes} \\
        \cmidrule[lr]{2-2}
        & ENISA Threat Landscape: List of top 15 threats~\cite{ref:enisa-threat-landscape-2022} \\
        \bottomrule
    \end{tblr}
    \caption{List of Documents Used to Identify the Attack Vectors.}
    \label{tab:list-of-documents-to-identify-attack-vectors}
\end{table}

According to the Verizon report, studying 18,419 cyberattacks, over 70\% were web-based attacks, 30\% were malware-based, and 20\% were social engineering attacks~\cite{ref:verizon-2022-data-breach-investigation}. Note that some of the attack vectors may overlap, resulting in the total percentage being over 100\%. The Ponemon report also shows the top three attack vectors faced by SMBs are phishing/social engineering, web-based attacks, and malware~\cite{ref:ponemon-2019-global-state-of-cybersecurity}. The Hiscox Cyber Readiness Report paints a similar picture on the frequent attack variables, labeling phishing as the most prevalent attack vector, followed by ransomware~\cite{ref:hiscox-cyber-readiness-2022}.

CISA labeled ransomware as the most visible cyberattack faced by businesses in the US~\cite{ref:chidukwani-a-survey-on-the-cyber-security-of-small-to-medium-businesses}. A report published by the ENISA states that phishing and malware attacks are the most common attacks faced by SMBs~\cite{ref:enisa-cybersecurity-for-smes}, and another report also published by ENISA declares malware, web-based attacks, and phishing to be the top three attacks~\cite{ref:enisa-list-of-top-15-threats}.

Based on the aforementioned reports, we decided to narrow the scope of our research to phishing/social engineering, malware/ransomeware, and web-based attacks. First, the three attacks cover a significant percentage of the attacks SMBs face. Second, SMBs do not have a significant workforce or budget to deal with all possible types of cyberattacks.

\subsection{Identifying Best Practices and Extracting Keywords}
\label{subsec:identifying-best-practices-and-extracting-keywords}
In order to obtain the desired set of TKSA for the defense against the three attack vectors, this paper uses a well-known keyword extractor, Yet Another Keyword Extractor! (YAKE!)~\cite{ref:campos-yake} on the documents published by government agencies and standardization institutions/organizations as shown in Table~\ref{tab:list-of-documents-categorized}.

\begin{table}[htbp]
    \centering
    \begin{tblr}{vspan=even, colspec={lX[l]c}, rowsep=2pt, width=\linewidth, rows={m}}
        \toprule
        Publisher & \SetCell[]{c}Document & Attack Vector \\
        \toprule
        \SetCell[r=4]{l}CISA & Capacity Enhancement Guide: Counter-phishing recommendations for Non-Federal Organizations~\cite{ref:cisa-counter-phishing-recommendations-for-non-federal} & \SetCell[r=4]{c}{\makecell{PSE \\ MR \\ WB}} \\
        \cmidrule[lr]{2-2}
        & Capacity Enhancement Guide: Counter-phishing recommendations for Federal Organizations~\cite{ref:cisa-counter-phishing-recommendations-for-federal} & \\
        \cmidrule[lr]{2-2}
        & CISA Website Security~\cite{ref:cisa-website-security} & \\
        \cmidrule[lr]{2-2}
        & CISA Ransomware Guide~\cite{ref:cisa-insights-ransomware-outbreak} & \\
        \midrule
        ACSC & ACSC Ransomware Prevention and Protection Guide~\cite{ref:acsc-ransomware-prevention-guide-2022} & MR \\
        \midrule
        NCSC & NCSC Mitigating Malware and Ransomware Attacks~\cite{ref:ncsc-mitigating-malware-ransomware} & MR \\
        \midrule
        \SetCell[r=2]{l}NIST & NIST SP 800-83r1: Guide to Malware & \SetCell[r=2]{c}MR \\
        \cmidrule[lr]{2-2}
        & Incident Prevention and Handling for
        Desktops and Laptops~\cite{ref:souppaya-guide-to-malware-nist} & \\
        \midrule
        ENISA & ENISA Threat Landscape 2020 – Web-based Attacks~\cite{ref:enisa-etl2020-web-based-attacks} & WB \\
        \midrule
        FTC & Cybersecurity for Small Businesses: Phishing~\cite{ref:ftc-cybersecurity-phishing} & PSE \\
        \midrule
        CISA & CISA Insights: Mitigations and hardening guidance from MSPs and Small and Mid-Sized Businesses~\cite{ref:cisa-guidance-msps-smbs} & MR \\
        \bottomrule
        \SetCell[c=3]{l}\footnotesize PSE = Phishing/Social Engineering \\
        \SetCell[c=3]{l}\footnotesize MR = Malware/Ransomware \\
        \SetCell[c=3]{l}\footnotesize WB = Web-Based Attack \\
    \end{tblr}
    \caption{List of Documents Categorized by Publisher and Attack Vector.}
    \label{tab:list-of-documents-categorized}
\end{table}

Keyword extraction has been widely used to derive knowledge maps~\cite{ref:chen-science-mapping, ref:yoon-development-and-application-of-a-keyword-based-knowledge-map-for-effective-rd-planning, ref:manesh-knowledge-management-in-the-fourth-industrial-revolution}. YAKE! allows us to achieve better results compared to other state-of-the-art keyword extraction methods such as Rake and TextRank~\cite{ref:campos-yake}. The n-gram parameter, which stands for the size of a sequence of terms in a keyword, was suggested to be set at 3 for best results~\cite{ref:campos-yake}.

To ensure the credibility and quality of the best practices, we restricted our pool of documents to limited but highly credible and well-established ones from organizations such as NIST, CISA, Federal Trade Commission (FTC), ENISA, National Cyber Security Centre (NCSC), and the Australian Cybersecurity Centre (ACSC) as shown in Table~\ref{tab:list-of-documents-categorized}. Table~\ref{tab:keyword-snippet} shows a snippet of keywords extracted from a document~\cite{ref:cisa-counter-phishing-recommendations-for-non-federal} along with their score (S). S is based on keyword features (term casing, term position, term frequency normalization, term relatedness to context, term different sentence) and is computed by the YAKE! algorithm. The lower the value of S, the more significant the keyword~\cite{ref:sun-a-review-of-unsupervised-keyphrase-extraction-methods-using-within-collection-resources}.

\begin{table}[htbp]
    \centering
    \begin{tblr}{vspan=even, colspec={X[l]cc}, rowsep=2pt, width=\linewidth, rows={m}}
        \toprule
        Keyword & Score (S) & After De-Duplication \\
        \toprule
        Secure Gateway Capabilities & 0.0023 & \\
        \midrule
        Stop Phishing Emails & 0.0090 & Removed \\
        \midrule
        Secure email gateways & 0.0102 & \\
        \midrule
        Gateway Capabilities & 0.0132 & Removed \\
        \midrule
        Secure Gateway & 0.0345 & Removed \\
        \midrule
        Gateways & 0.1013 & \\
        \midrule
        Email filter solution & 0.1434 & \\
        \midrule
        Signatures and blocklists & 0.1481 & Removed \\
        \midrule
        Host Level Protections & 0.1710 & \\
        \bottomrule
    \end{tblr}
    \caption{Snippet of Keywords Extracted from~\cite{ref:cisa-counter-phishing-recommendations-for-non-federal}.}
    \label{tab:keyword-snippet}
\end{table}

To eliminate similar keywords, we employed a de-duplication process based on similarity algorithms such as Levenshtein similarity~\cite{ref:levenshtein-binary-codes-capable-of-correcting-deletions-insertions-and-reversals}, Jaro-Winkler~\cite{ref:winkler-string-comparator-metrics-and-enhanced-decision-rules-in-the-fellegi-sunter-model-of-record-linkage}, and Hamming Distance~\cite{ref:hamming-error-detecting-and-error-correcting-codes, ref:prasetya-performance-of-text-similarity-algorithms}. We used Levenshtein similarity because it works on the principle of the minimum number of single-character edits required to change one word into the other~\cite{ref:prasetya-performance-of-text-similarity-algorithms}. For example, take a group of similar keywords like \textit{Secure Gateway Capabilities}, \textit{Gateway Capabilities}, and \textit{Secure Gateway}, all of which are similar and can be considered as just one keyword, \textit{Secure Gateway Capabilities}. Table~\ref{tab:keyword-snippet} shows a snippet of keywords after the de-duplication process is applied, the \textit{After De-Duplication} column shows if the keyword was removed after the de-duplication process.

\subsection{Mapping Keywords to The NICE Framework}
\label{subsec:mapping-keywords-to-the-nice-framework}
\begin{figure}
    \centering
    \includegraphics[max width=0.5\textwidth]{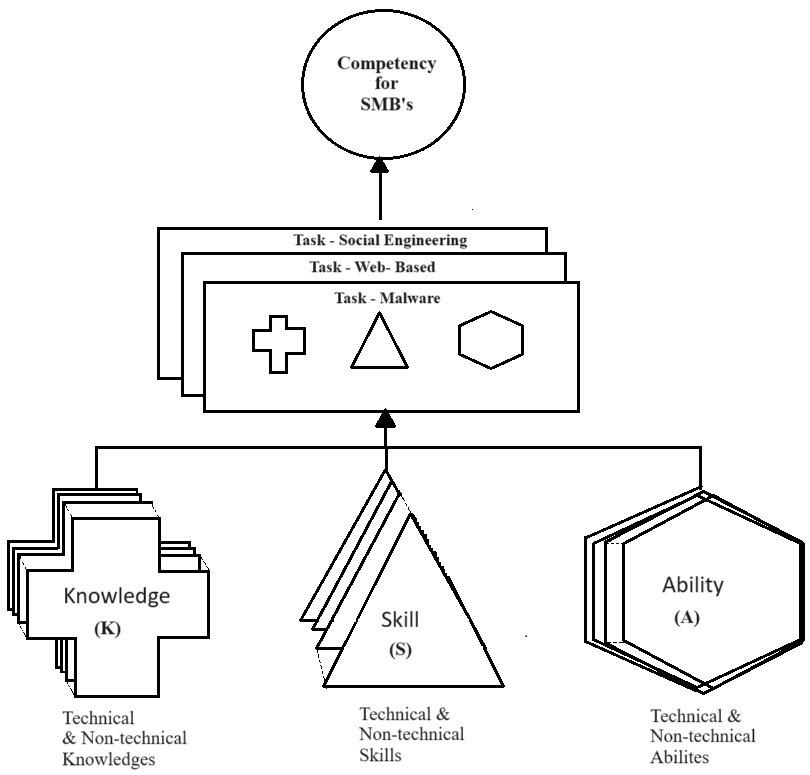}
    \caption{Relationship Between Competency and TKSA.}
    \label{fig:competency-tksa-relationship}
\end{figure}

As shown in Figure~\ref{fig:competency-tksa-relationship} our goal is to map TKSAs to competencies related to the attack vectors relevant to SMBs identified above. For convenience, we divided the result of the mapping exercise into two models, technical and non-technical. The technical model consists of all the TKSA that involve a certain level of proficiency in the technical aspect of cybersecurity, whereas the non-technical model consists of TKSA that are related to general cyber awareness, legal, and managerial proficiencies. Most of the non-technical TKSA can be applied to all employees of an SMB. Sections~\ref{subsubsec:technical-model} and~\ref{subsubsec:non-technical-model} explain the technical and non-technical model respectively.

\subsubsection{Technical Model}
\label{subsubsec:technical-model}
We mapped 49 technical knowledges covering 7.57\% of all the knowledges in the NICE Framework, 23 technical skills covering 6.10\% of all the skills in the NICE Framework, 6 technical abilities covering 3.38\% of all the abilities in the NICE Framework, and 10 technical tasks covering 1.02\% of all the tasks in the NICE Framework, totaling 88 technical TKSA. These TKSA all address a different aspect of dealing with the three attack vectors. Table~\ref{tab:technical-knowledge-mapped} depicts the technical knowledges mapped to the three different attack vectors. K0002 for example, \textit{knowledge of risk management processes (e.g., methods for assessing and mitigating risk)}; is a broad statement that can be used in all three attack cases. Conversely, K0105---\textit{knowledge of web services (e.g., service-oriented architecture, Simple Object Access Protocol, and web service description language)}---and K0188---\textit{knowledge of malware analysis tools (e.g., Oily Debug, Ida Pro)}---are specific to an attack.

\begin{table}[htbp]
    \centering
    \begin{tblr}{vspan=even, colspec={lX[l]ccc}, rowsep=2pt, width=\linewidth, rows={m}}
        \toprule
        \makecell{TKSA \\ Number} & \SetCell[]{c}TKSA Description & PSE & MR & WB \\
        \toprule
        K0001 & Knowledge of computer networking concepts and protocols, and network security methodologies. & & & * \\
        \midrule
        K0002 & Knowledge of risk management processes (e.g., methods for assessing and mitigating risk). & * & * & * \\
        \midrule
        K0004 & Knowledge of cybersecurity and privacy principles. & * & * & * \\
        \midrule
        K0005 & Knowledge of cyber threats and vulnerabilities. & * & * & * \\
        \midrule
        K0007 & Knowledge of authentication, authorization, and access control methods. & * & * & * \\
        \midrule
        K0013 & Knowledge of cyber defense and vulnerability assessment tools and their capabilities. & * & * & * \\
        \midrule
        K0033 & Knowledge of host/network access control mechanisms (e.g., access control list, capabilities lists). & & & * \\
        \midrule
        K0038 & Knowledge of cybersecurity and privacy principles used to manage risks related to the use, processing, storage, and transmission of information or data. & & * & \\
        \bottomrule
        \SetCell[c=3]{l}\footnotesize PSE = Phishing/Social Engineering \\
        \SetCell[c=3]{l}\footnotesize MR = Malware/Ransomware \\
        \SetCell[c=3]{l}\footnotesize WB = Web-Based Attack \\
    \end{tblr}
    \caption{Snippet of technical knowledge mapped to the three attack vectors (7.8\% of all the knowledge in the NICE Framework)}
    \label{tab:technical-knowledge-mapped}
\end{table}

56.26\% of the technical knowledges we mapped apply to just one specific attack vector; i.e., they cannot be applied to a different attack than the ones they are mapped to. 16.66\% of the technical knowledges are mapped to two attacks; i.e., it can vary based on the description of the knowledge and they can be used in two specific attacks out of the three. For example, K0202 applies to phishing/social engineering attacks and malware/ransomware attacks but does not apply to web-based attacks. The remaining 27.08\% of the knowledges apply to all three attack vectors.

The skills and abilities are the least versatile among the technical TKSAs, with 56.53\% of the technical skills and 83.33\% of the technical abilities applying to only one specific attack vector. The knowledges are to be the most versatile among the technical TKSA with 40\% of them applying to all three attack vectors, and 50\% applying to at least two specific attack vectors.

\subsubsection{Non-Technical Model}
\label{subsubsec:non-technical-model}
The non-technical model consists of 28 non-technical knowledges covering 4.41\% of all the knowledges in the NICE Framework, 8 non-technical skills covering 2.12\% of all the skills in the NICE Framework, 7 non-technical abilities covering 3.95\% of all the abilities in the NICE Framework, and 11 non-technical tasks covering 1.12\% of all the tasks in the NICE Framework, adding up to a total of 54 non-technical TKSA. They paint a different picture when compared to the technical model due to most of the general non-technical TKSA being applicable to all the 3 attack vectors. Table~\ref{tab:non-technical-knowledge-mapped} depicts how non-technical knowledges are mapped to the three attack vectors. 

\begin{table}[htbp]
    \centering
    \begin{tblr}{vspan=even, colspec={lX[l]ccc}, rowsep=2pt, width=\linewidth, rows={m}}
        \toprule
        \makecell{TKSA \\ Number} & \SetCell[]{c}TKSA Description & PSE & MR & WB \\
        \toprule
        K0003 & Knowledge of laws, regulations, policies, and ethics as they relate to cybersecurity and privacy. & * & * & * \\
        \midrule
        K0006 & Knowledge of specific operational impacts of cybersecurity lapses. & * & * & * \\
        \midrule
        K0066 & Knowledge of Privacy Impact Assessments. & * & * & * \\
        \midrule
        K0098 & Knowledge of the cyber defense Service Provider reporting structure and processes within one’s own organization. & * & * & * \\
        \midrule
        K0101 & Knowledge of the organization’s enterprise information technology (IT) goals and objectives. & * & * & * \\
        \midrule
        K0107 & Knowledge of Insider Threat investigations, reporting, investigative tools and laws/regulations. & * & * & * \\
        \midrule
        K0123 & Knowledge of legal governance related to admissibility (e.g. Rules of Evidence). & * & * & * \\
        \midrule
        K0126 & Knowledge of Supply Chain Risk Management Practices (NIST SP 800-161) & & * & \\
        \bottomrule
        \SetCell[c=3]{l}\footnotesize PSE = Phishing/Social Engineering \\
        \SetCell[c=3]{l}\footnotesize MR = Malware/Ransomware \\
        \SetCell[c=3]{l}\footnotesize WB = Web-Based Attack \\
    \end{tblr}
    \caption{Snippet of non-technical knowledge mapped to the three attack vectors (4.4\% of all the knowledge in the NICE Framework)}
    \label{tab:non-technical-knowledge-mapped}
\end{table}

93\% of the non-technical knowledges apply to all 3 attack vectors, with only 7\% of the non-technical knowledges applying to one specific attack. 85\% of the non-technical abilities and 87.5\% of the non-technical skills apply to all 3 attack vectors, while 91\% of the non-technical tasks apply to all 3 attack vectors.

When both models are compared, we noticed the non-technical model to be more versatile than the technical model. This is in line with our expectations. The technical model proves to be more technically sound and specific to particular attack vectors while the non-technical model applies to cybersecurity in general and is not heavily based on attack vectors.

\subsection{Integration of KSAs, Virtual Machine Labs, and NICE Framework}
\label{subsec:integration-of-KSAs-virtual-machine-labs-and-NICE-framework}
A robust cybersecurity and legal education strategy integrates legal frameworks/regulations and KSAs with hands-on practice in virtual machine labs, guided by the NICE framework. This ensures that learners not only grasp theoretical concepts but also acquire practical skills to tackle real-world cybersecurity and legal challenges. The NICE Framework's structured roles and competencies guide curriculum development, aligning programs with industry standards. Virtual machine labs provide a practical setting for learners to apply theoretical knowledge, simulate cyberattacks, and build skills in areas such as network security and incident response. Each lab aligns with specific KSAs and is augmented by legal case studies and frameworks, reinforcing learning and building confidence in managing cybersecurity issues. This integrated approach enhances employability, adapts to emerging threats, and ensures that education programs remain relevant and effective.

This research introduces a tailored legal and cybersecurity curriculum designed specifically to meet the needs of SMBs, with a focus on the most relevant threats such as phishing/social engineering, malware/ransomware, and web-based attacks (Objective 1). Through a scenario-based learning model (Objective 3), students participate in realistic threat simulations, enhancing their problem-solving and practical skills in identifying and mitigating these targeted risks~\cite{ref:bada-2019-cybersecurity-awareness-campaigns}. This curriculum, therefore, prepares both law and computer science (CS) students for roles involving cybersecurity risk assessment, regulatory compliance, and legal forensics. Key scenarios include managing malware (e.g., Eternal Blue and Spectre/Meltdown), responding to web-based attacks (e.g., PBX Hacking and Website Fingerprinting), and understanding social engineering tactics (e.g., DDoS and Hacking Group Thallium), providing hands-on experience that prepares students for real world challenges~\cite{ref:nice-cybersecurity-workforce-framework}.

The curriculum also emphasizes the development of practical skills using virtual machine labs through the Ohio Cyber Range Institute (OCRI)~\cite{ref:ocri} resources (Objective 4), offering immersive training in cybersecurity roles. Aligned with the NICE Framework, it fosters the development of essential KSAs~\cite{ref:shojaifar-automating-communication-of-cybersecurity}, integrating both technical and legal knowledge (Objective 2) to equip students with a comprehensive understanding of cybersecurity~\cite{ref:shojaifar-automating-communication-of-cybersecurity}. This holistic approach ensures graduates are well-prepared to defend SMBs against evolving cyber threats and comply to legal regulations, bridging the gap between theory and practice in cybersecurity education.

\section{Scenario-Based Curriculum}
\label{subsec:scenario-based-curriculum}

A scenario-based model in cybersecurity education uses realistic situations to teach and illustrate cybersecurity concepts and pairs them with appropriate legal frameworks, regulations, and case studies. Learners engage in scenarios that simulate actual cyber threats, e.g., phishing attacks, malware infections, or data breaches, to understand and apply cybersecurity principles in practice~\cite{ref:ghosh-assessing-competencies-using-scenario-based-learning-in-cybersecurity}.

In this curriculum, learners analyze scenarios, identify relevant cybersecurity concepts and regulations, and develop appropriate responses, some of which may be legally required. This method enhances theoretical understanding by demonstrating practical applications and integrating legal considerations---such as data privacy laws and regulatory compliance---into the scenarios~\cite{ref:BRILINGAITE2020101607}.

The curriculum features six key scenarios targeting malware/ransomware, web-based attacks, and phishing/social engineering, each designed to provide practical experience:

\begin{description}
    \item[Scenario 1:] Eternal Blue---focuses on a malware attack exploiting the Eternal Blue vulnerability, emphasizing patch management.
    \item[Scenario 2:] Spectre/Meltdown---addresses hardware-level vulnerabilities and the need for robust hardware security.
    \item[Scenario 3:] PBX Hacking---deals with web-based attacks on private branch exchange (PBX) systems, highlighting network security challenges.
    \item[Scenario 4:] Website Fingerprinting---explores attacks through website fingerprinting, stressing web traffic protection.
    \item[Scenario 5:] DDoS Attack---involves a distributed denial of service (DDoS) attack, focusing on network resilience.
    \item[Scenario 6:] Hacking Group Thallium---examines a phishing attack by the Thallium group, emphasizing defense against social engineering.
\end{description}

Each scenario is paired with hands-on labs using OCRI-supported virtual machines and legal materials, ensuring a thorough and practical approach to skill development. The scenarios are explained in further detail below in Section~\ref{subsec:overview-of-scenarios} and the course syllabus is given in Table~\ref{tab:course-syllabus}.

\begin{longtblr}[
    caption = {Course Syllabus.},
    label = {tab:course-syllabus}
]{vspan=even, colspec={p{1.0cm}p{2.5cm}p{2.5cm}X[l]X[l]}, rowsep=2pt, width=\linewidth, rows={m}, rowhead=1}
    \toprule
    Week & Topic & Lab & Readings & Videos/Assignment/Quiz \\
    \toprule
    
    1 & 
    Introduction & 
    & 
    \textit{Global Investigations Review, The Guide to Cybersecurity Investigations} (2nd ed. 2021), \href{https://globalinvestigationsreview.com/guide/the-guide-cyber-investigations-archived/first-edition/article/the-cyber-threat-landscape}{Ch. 1: \textit{The Threat Landscape}}
    \vspace{5pt}\newline 
    Robert M. Chesney \textit{\href{https://papers.ssrn.com/sol3/papers.cfm?abstract_id=3547103}{Cybersecurity Law, Policy, and Institutions (v. 3.1)}}---Introduction to Key Terms \& 
    Concepts---pp. 11-16 & 
    \textit{\href{https://www.youtube.com/watch?v=N20q-ZMop0w}{Russia vs. Ukraine: The Biggest Cyber Attack Ever. Darknet Diaries Ep. 54: NotPetya}} \\

    \midrule

    2 &
    Ransomware \newline Investigations &
    Eternal Blue &
    The Sedona Conference, \href{https://thesedonaconference.org/publication/Incident_Response_Guide}{\textit{Incident Response Guide}} (2020), pp. 133-150 &
    \href{https://www.youtube.com/watch?v=PKHH_gvJ_hA}{\textit{WANNACRY: The World’s Largest Ransomware Attack}} \\

    \midrule

    3 &
    Ransomware \newline Investigations &
    Eternal Blue &
    \href{https://www.morganlewis.com/pubs/2017/05/wannacry-ransomware-cyberattack-raises-legal-issues}{\textit{WannaCry Ransomware CyberAttack Raises Legal Issues}} (May 22, 2017)
    \vspace{5pt}\newline
    \textbf{Optional Background Reading:} Michael Schmitt and Sean Fahey, \href{https://www.justsecurity.org/50038/wannacry-international-law-cyberspace/}{\textit{WannaCry and the International Law of Cyberspace}} (Dec. 22, 2017) &
    \href{https://www.rsaconference.com/library/Presentation/USA/2021/hooked-by-phisherman-quarterbacking-breach-response-with-law-enforcement}{\textit{Hooked By Phisherman: Quarterbacking Incident Response with Law Enforcement Summary}}
    \vspace{5pt}\newline
    Eternal Blue Lab Report Due
    \vspace{5pt}\newline
    Quiz 1 \\

    \midrule

    4 &
    Data Breach \newline Notification \newline Laws &
    Meltdown/Spectre & 
    \textit{Global Investigations Review, The Guide to Cybersecurity Investigations} (3rd. ed. 2023) \href{https://globalinvestigationsreview.com/guide/the-guide-cyber-investigations/third-edition/article/regulatory-compliance-in-the-context-of-cross-border-data-breach}{Ch. 4: \textit{Regulatory Compliance in the Context of a Cross-Border Data Breach}}
    \vspace{5pt}\newline
    BakerHostetler \textit{US Data Breach Notification Law Interactive Map}---read the specific requirements in the section following the map &
    \href{https://www.youtube.com/watch?v=lQZzm9z8g_U}{\textit{Keynote: Spectre, Meltdown, \& Linux---Greg Kroah-Hartman, Fellow, The Linux Foundation}}
    \vspace{5pt}\newline
    \textit{Data Breach Notification Laws} \\

    \midrule

    5 & 
    Data Breach \newline Notification \newline Laws &
    Meltdown/Spectre &
    Aaron Simpson and Adam H. Solomon, \href{https://www.huntonak.com/images/content/5/8/v2/58221/4_Complying_with_Breach_Notification_Obligations.pdf}{\textit{Complying with Breach Notification Obligations in a Global Setting: A Legal Perspective}} &
    Meltdown/Spectre Lab Report 
    \vspace{5pt}\newline
    Quiz 2 \\

    \midrule

    6 &
    Cybersecurity \newline Risk \newline Assessments & 
    PBX Hacking &
    \href{https://www.celerium.com/cybersecurity-frameworks-a-comprehensive-guide}{\textit{The Complete Guide to Understanding Cybersecurity Frameworks}}
    \vspace{5pt}\newline
    Practical Law \textit{Data Security Risk Assessments and Reporting} (2021) &
    \textit{Cybersecurity Risk Assessment} \\

    \midrule

    7 &
    Cybersecurity \newline Risk \newline Assessments &
    PBX Hacking &
    \textit{Performing Data Security Risk Assessments Checklist}
    \vspace{5pt}\newline
    \href{https://csulawonline.instructure.com/courses/193/files/8623?wrap=1}{Cybersecurity Tech Basics: Critical Security Controls: Overview } &
    PBX Hacking Lab Report 
    \vspace{5pt}\newline
    Quiz 3 \\

    \midrule

    8 &
    International Law Issues &
    Website \newline Fingerprinting &
    Cyberlaw CCDCOE, \href{https://cyberlaw.ccdcoe.org/wiki/Scenario_24:_Internet_blockage}{\textit{Scenario 24: Internet Blockage 1.1-2.1}} &
    RSAC, \textit{Cyber and Modern Conflict: The Changing Face of Modern Warfare} (June 7, 2022) \\

    \midrule

    9 & 
    \SetCell[c=4]{halign=c}\centering\textbf{Spring Break}\par \\

    \midrule

    10 &
    International Law Issues &
    Website \newline Fingerprinting &
    Cyberlaw CCDCOE, \href{https://cyberlaw.ccdcoe.org/wiki/Scenario_24:_Internet_blockage}{\textit{Scenario 24: Internet Blockage 1.1-3}} &
    Website Fingerprinting Lab Report
    \vspace{5pt}\newline
    Quiz 4 \\

    \midrule

    11 &
    Criminal Law &
    \#OPJustina: DDoS &
    David Kushner, \href{https://www.rollingstone.com/culture/culture-features/the-hacker-who-cared-too-much-196425/}{\textit{The Attacker Who Cared Too Much}} (June 29, 2017) &
    Darknet Diaries \href{https://www.youtube.com/watch?v=0qvBYj7F3jo&list=PLtN43kak3fFEEDNo0ks9QVKYfQpT2yUEo&index=14}{Ep. 14: \textit{OpJustina}} \\

    \midrule

    12 &
    Criminal Law &
    \#OPJustina: DDoS &
    Chesney on Cybersecurity---\textit{Computer Fraud \& Abuse Act}---pp. 17-30 &
    \href{https://www.youtube.com/watch?v=cPr6hZfoBfQ}{\textit{US Computer Fraud and Abuse Act (CFAA)}}
    \vspace{5pt}\newline
    \#OPJustina DDoS Lab Report
    \vspace{5pt}\newline
    Quiz 5 \\

    \midrule

    13 &
    Nation-State \par Cybercrime &
    Thallium &
    Tom Burt \href{https://blogs.microsoft.com/on-the-issues/2019/12/30/microsoft-court-action-against-nation-state-cybercrime/}{\textit{Microsoft takes court action against fourth nation-state cybercrime group}}
    \vspace{5pt}\newline
    \href{https://noticeofpleadings.com/thallium/}{Thallium Civil Case Documents} &
    CyberWire, \href{https://www.youtube.com/watch?v=jGC-UwKVkYc}{Taking down Thallium} \\

    \midrule

    14 &
    Nation-State \par Cybercrime &
    Thallium &
    Lubin \& Marinotti, \href{https://www.lawfaremedia.org/article/why-current-botnet-takedown-jurisprudence-should-not-be-replicated}{\textit{Why Current Botnet Takedown Jurisprudence Should Not Be Replicated}} &
    Thallium Lab Report 
    \vspace{5pt}\newline
    Quiz 6 \\

    \bottomrule
\end{longtblr}

\subsection{Scenarios and TKSAs}
\label{subsec:scenarios-and-tksas}
We have identified 49 technical knowledge areas and 29 non-technical knowledge areas which are essential for protecting SMBs. Tables~\ref{tab:non-technical-tksa-mapping-to-vectors} and ~\ref{tab:technical-tksa-mapping-to-vectors} below map these TKSAs to the six key scenarios, organized by category. This categorization simplifies the identification of required TKSAs and underscores the specific areas of expertise needed to address each type of threat.

\begin{table}[htbp]
    \centering
    \begin{tblr}{vspan=even, colspec={ccX[l]}, rowsep=2pt, width=\linewidth, rows={m}}
        \toprule
        \makecell{Attack \\ Vector} & Scenarios & \SetCell[]{c}Non-Technical TKSAs \\
        \toprule
        MR & \makecell{Scenario 1 \\ Scenario 2} & K0003, K0006, K0098, K0101, K0123, K0150, K0264, K0287, K0315, K0351, K0429, K0504, K0511, K0524, K0585, A0009, A0033, A0046, A0110, A0113, A0115, A0146 \\
        \midrule
        WB & \makecell{Scenario 3 \\ Scenario 4} & K0003, K0006, K0098, K0101, K0123, K0150, K0264, K0287, K0315, K0351, K0429, K0504, K0511, K0524, K0585, A0033, A0046, A0110, A0113, A0115, A0146 \\
        \midrule
        PSE & \makecell{Scenario 5 \\ Scenario 6} & K0003, K0006, K0098, K0101, K0123, K0150, K0264, K0287, K0315, K0351, K0429, K0504, K0511, K0524, K0585, S0085, S0213, S0219, S0232, S0361, T0099, T0280, A0033, A0046, A0110, A0113, A0115, A0146 \\
        \bottomrule
        \SetCell[c=3]{l}\footnotesize PSE = Phishing/Social Engineering \\
        \SetCell[c=3]{l}\footnotesize MR = Malware/Ransomware \\
        \SetCell[c=3]{l}\footnotesize WB = Web-Based Attack \\
    \end{tblr}
    \caption{List of Non-Technical TKSA’s Mapped to the Three Attack Vectors.}
    \label{tab:non-technical-tksa-mapping-to-vectors}
\end{table}
    
\begin{table}[htbp]
    \centering
    \begin{tblr}{vspan=even, colspec={ccX[l]}, rowsep=2pt, width=\linewidth, rows={m}}
        \toprule
        \makecell{Attack \\ Vector} & Scenarios & \SetCell[]{c}Technical TKSAs \\
        \toprule
        MR & \makecell{Scenario 1 \\ Scenario 2} & K0002, K0004, K0005, K0007, K0013, K0038, K0046, K0049, K0070, K0073, K0074, K0104, K0135, K0176, K0188, K0189, K0205, K0210, K0260, K0261, K0274, K0368, K0392, K0480, K0516, K0536, K0624, K0626, K0627, S0001, S0022, S0076, S0084, S0121, S0192, S0264, S0298, T0056, T0161, T0181, T0271, T0438, T0553, T0751, A0010, A0062 \\
        \midrule
        WB & \makecell{Scenario 3 \\ Scenario 4} & K0001, K0002, K0004, K0005, K0013, K0033, K0046, K0049, K0056, K0058, K0062, K0070, K0100, K0104, K0131, K0176, K0189, K0202, K0332, K0392, K0427, K0452, S0001, S0004, S0022, S0046, S0076, S0084, S0121, S0192, S0258, S0264, T0023, T0161, T0271, T0438, T0553, A0176 \\
        \midrule
        PSE & \makecell{Scenario 5 \\ Scenario 6} & K0002, K0004, K0005, K0007, K0013, K0046, K0049, K0104, K0131, K0202, K0205, K0210, K0260, K0274, K0336, K0368, K0392, K0452, S0022, S0076, S0084, S0121, S0192, T0056, T0271, T0438, T0553, A0063, A0119, A0123, A0176 \\
        \bottomrule
        \SetCell[c=3]{l}\footnotesize PSE = Phishing/Social Engineering \\
        \SetCell[c=3]{l}\footnotesize MR = Malware/Ransomware \\
        \SetCell[c=3]{l}\footnotesize WB = Web-Based Attack \\
    \end{tblr}
    \caption{List of Technical TKSAs Mapped to the Three Attack Vectors.}
    \label{tab:technical-tksa-mapping-to-vectors}
\end{table}

\subsubsection{Leveraging Virtual Machine Labs}
\label{subsubsec:levaraging-virtual-machine-labs}
Virtual machine labs play a crucial role in developing cybersecurity skills by providing an interactive, hands-on learning environment. These labs simulate real-world cyber threats, allowing learners to enhance their skills in a controlled setting.

Our VM lab uses the OCRI platform, which hosts virtual machines such as Kali Linux, Windows 9, and Windows 11, offering a comprehensive environment for cybersecurity training. OCRI supports advanced education through scalable and accessible labs that replicate real-world attack scenarios.

The flexibility of OCRI labs allows learners to access them remotely, catering to diverse schedules and skill levels. This approach ensures practical learning experiences, preparing students to effectively handle cybersecurity challenges.

\subsubsection{OCRI Exercises}
\label{subsubsec:ocri-exercises}
The OCRI educational resources include 14 modules covering topics like cryptography, digital forensics, network security, and encryption, with exercises designed to enhance practical cybersecurity skills. Our curriculum incorporates selected exercises from these modules to provide hands-on experience with specific threat vectors. Modules such as \textit{Introduction to Digital Forensics} and \textit{Defense in Depth Network Security} are used to teach essential skills in areas such as malware analysis, network and firewall configuration, password security, and encryption. These exercises offer students practical, scenario-based learning opportunities, bridging the gap between theoretical knowledge and real-world cybersecurity challenges.

\subsubsection{Legal Considerations}
\label{subsubsec:legal-considerations}
Understanding the legal landscape is critical in cybersecurity as it helps organizations comply with regulations, handle cyber incidents, and manage data breaches. Each of the scenarios is paired with several key legal readings and resources to augment the lesson and demonstrate the intersection of legal requirements and cybersecurity. These materials cover important topics such as ransomware investigations, data breach notification laws and other regulatory compliance, and international legal issues. By integrating legal insights throughout the curriculum, learners gain a comprehensive understanding of how to navigate the legal complexities that accompany cybersecurity challenges.

\subsection{Overview of Scenarios}
\label{subsec:overview-of-scenarios}
Below, we give a brief overview of each of the scenarios used in the curriculum, including the hands-on lab and legal resources. Table~\ref{tab:scenarios-description} gives an overview of each scenario including a description and the integrated non-technical skills.

\begin{longtblr}[
    caption = {Cybersecurity Scenarios with Attacks and Non-Technical Topics.},
    label = {tab:scenarios-description}
]{vspan=even, colspec={p{2.5cm}X[l]X[l]}, rowsep=2pt, width=\linewidth, rows={m}, rowhead=1}
    \toprule
    Scenarios & Scenario Description (Part 1) & Topics Illustrated (Part 2) \\
    \toprule

    \#1 EternalBlue \par (Week 2-3)
    & Ransomware attack exploited the vulnerabilities in Windows OS, affecting online services of numerous cities including Baltimore (2019). Earlier, NSA discovered the vulnerabilities but did not report to Microsoft. Experiment with the VirtualBox environment.
    & \underline{Non-technical topics}: Data breach notification laws; identifying technical “triggers” for reporting; role of law enforcement; data privacy laws. 
    \underline{Technical topics}: Basic security concepts (confidentiality, integrity, and availability); Encryption; Malware (Virus, Trojan); Hardening OS.
    \\ \midrule

    \#2 Meltdown/Spectre \par (Week 4-5)
    & In 2015, Google's Project Zero team discovered a hardware vulnerability, affecting a wide range of systems running iOS, Linux, macOS, and Windows. Software workaround has been assessed as slowing computers between 5$\sim$30\%, although companies responsible for software correction of the exploit are reporting minimal impact from general benchmark testing.
    & \underline{Non-technical topics}: Legal responsibility/liability for hardware vulnerabilities; legal analysis of when to disclose vulnerabilities. 
    \underline{Technical topics}: Hardware security (hardware trojan, reverse engineering, side channel attacks); IoT security.
    \\ \midrule

    \#3 PBX Hacking \par (Week 6-7)
    & A cyber-criminal, who is on the Interpol and the FBI's most wanted list, was arrested (2015). He let other people's phones dialed pay-per-minute numbers that they owned, making money. Experiment with FreePBX system inside VirtualBox.
    & \underline{Non-technical topics}: Challenges in addressing low-level cybercrime; limits of law enforcement. 
    \underline{Technical topics}:  Access control (firewalls, ports, authentication, etc.), Voice over IP (VoIP).
    \\ \midrule

    \#4 Website Fingerprinting Attack \par (Week 8-9)
    & Internet censorship is on the rise. It allows a state agency to block the information flow to and from a black-listed websites. Experiment to show the distinguishable traffic pattern when visiting websites.
    & \underline{Non-technical topics}: Internet censorship laws; legal and ethical issues in compliance. 
    \underline{Technical topics}:  Computer network protocols including HTTP; VPN; Anonymous networks (Tor); Machine learning/AI.
    \\ \midrule

    \#5 \#OpJustina: DDoS Attack \par (Week 10-11)
    & DDoS attack on Boston Children's Hospital (2013) disrupting the hospital's day-to-day operations. Attacker gains control of a network of online machines and directs them to send requests to the target. Experiment with Mininet tesbed within VirtualBox environment. 
    & \underline{Non-technical topics}: Computer Fraud and Abuse Act; legal reform to address cybercrime. 
    \underline{Technical topics}:  Network security concepts; Anomaly detection.
    \\ \midrule

    \#6 Hacking Group Thallium \par (Week 12-13)
    & The US District Court for the Eastern District of Virginia unsealed documents related to a law suit filed by Microsoft against the cybercrime group Thallium (2019). It used malware to compromise systems and steal data, then create fake email addresses to launch phishing attacks to individual employees of the organization. 
    & \underline{Non-technical topics}: Cybercrime laws; practical and legal challenges in attribution; laws and procedures for shutting down servers used for cybercrime. 
    \underline{Technical topics}:  Ethical hacking; Phishing attacks; Identity theft; Social engineering.
    \\ \bottomrule
\end{longtblr}

\subsubsection{Scenario 1: Eternal Blue}
\label{subsubsec:scenario-1-eternal-blue}

\begin{figure}
    \centering
    \includegraphics[width=0.5\textwidth]{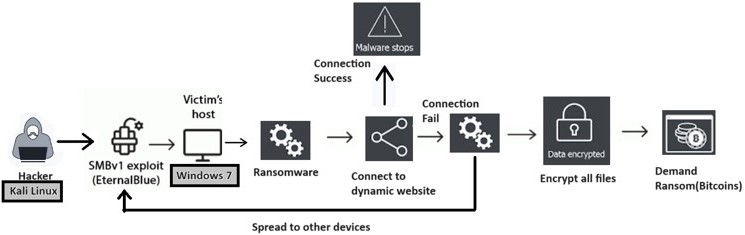}
    \caption{Eternal Blue Attack.}
    \label{fig:scenario-1}
\end{figure}

In this lab, participants engage in the simulation of the Eternal Blue attack~\cite{ref:eternal-blue}, a notorious exploit that capitalizes on the vulnerability within the SMBv1 protocol to propagate malware across networks. Figure~\ref{fig:scenario-1} provides a graphical overview of the scenario. Utilizing Kali Linux as the attacker’s platform and Windows 7 as the target system, the exercise begins with the acquisition and unpacking of ransomware from a specified URL, laying the groundwork for subsequent malicious activities. Leveraging the inherent weaknesses in Microsoft Window’s handling of specially crafted packets, participants exploit the Eternal Blue vulnerability to establish a meterpreter session, gaining unauthorized access to the victim’s machine. Subsequently, the exploit is employed to transmit the ransomware to the victim’s system, initiating the encryption process for all accessible files.

This scenario begins students' introduction to incident response. With the readings assigned for this scenario, students learn critical components of an incident response, including background, planning, and execution~\cite{ref:sedona-incident-response-guide}. The readings reinforce that a security incident is often considered inevitable and that planning for such an incident is critical and must be cross-functional~\cite{ref:sedona-incident-response-guide}. Building off a cross-functional incident response, legal obligations, such as the Health Insurance Portability and Accountability Act of 1996 (HIPAA)~\cite{ref:hipaa} and data breach notification laws, are incorporated into this scenario.

\subsubsection{Scenario 2: Spectre/Meltdown}
\label{subsubsec:scenario-2-eternal-blue}
This scenario involves students comprehending a spectre attack~\cite{ref:spectre-and-meltdown} on a Kali Linux machine. Additionally, students scrutinize legal issues and risks arising from hardware vulnerabilities in the context of cybersecurity practice, including laws that govern the responsibility for identifying vulnerabilities and potential liability for organizations affected by related cyberattacks. They will also learn to advise clients on vulnerability disclosure~\cite{ref:complying-with-breach-notification-obligations, ref:guide-to-cyber-investigations}.

\subsubsection{Scenario 3: PBX Hacking}
\label{subsubsec:scenario-3-pbx-hacking}
\begin{figure}
    \centering
    \includegraphics[width=0.5\textwidth]{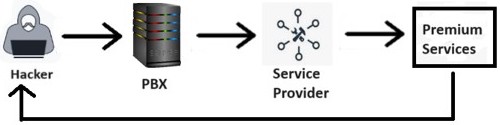}
    \caption{PBX Attack. Cyber Criminals exploit vulnerabilities to reroute calls to pay-per-minute lines, incurring substantial charges for the user.}
    \label{fig:scenario-3}
\end{figure}

As shown in Figure~\ref{fig:scenario-3}, this lab experiment explores security vulnerabilities in PBX administrative systems by focusing on the risk posed by weak passwords~\cite{ref:mcinnes-the-voip-pbx-honeypot-advance-persistent-threat-analysis, ref:khan-design-and-configuration-of-voip}. While threats like firewall bypasses and cross-site scripting (XSS) vulnerabilities in Free PBX exist, the primary goal is to emphasize the dangers of inadequate password security. The setup involves three machines: a Linux machine as the PBX hub, a Windows machine for PBX configuration, and a Kali Linux machine for probing the system. Using Kali Linux, the attacker intercepts HTTP requests between the PBX and Windows and employs Hydra, a password-cracking tool, to access the admin panel, highlighting the need for strong password protections.

The readings assigned for this scenario introduce students to the Computer Fraud and Abuse Act (CFAA)~\cite{ref:cfaa}. Through case studies where individuals were charged with violating the CFAA~\cite{ref:chesney-cybersecurity-law-policy-and-institutions}, students gain a fundamental understanding of what constitutes a cyber crime and the legal implications that follow. Other concepts discussed in this scenario are wire fraud~\cite{ref:wire-fraud} and identity fraud~\cite{ref:identity-fraud}.

\subsubsection{Scenario 4: Website Fingerprinting}
\label{subsubsec:scenario-4-pbx-hacking}
\begin{figure}
    \centering
    \includegraphics[width=0.5\textwidth]{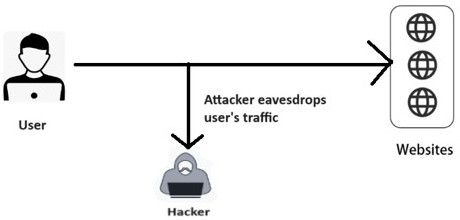}
    \caption{Website Fingerprinting Attack. Encrypted web traffic patterns are analyzed to infer user-visited websites and compromise privacy.}
    \label{fig:scenario-4}
\end{figure}

In this lab, students simulate a website fingerprinting attack using k Nearest Neighbors (kNN) on a Kali Linux machine. The goal is to analyze traffic patterns such as packet lengths, frequency, and timing to identify visits to blacklisted websites~\cite{ref:cherubin-online-website-fingerprinting}. Students conduct 15 browsing sessions for each site, capturing traffic data from seven websites using the Lynx browser. The data is generated by a tcpdump~\cite{ref:tcpdump} for each website visit and used to train a kNN classifier to determine whether a user is accessing any of the blacklisted websites by analyzing distinct patterns of web traffic. Figure~\ref{fig:scenario-4} shows the setup for this scenario. A case study from the NATO Cooperative Cyber Defence Centre of Excellence (CCDCE) reinforces this lesson by giving a real-world example of censorship and discussing the right to secure Internet access and privacy~\cite{ref:internet-blockage}.

\subsubsection{Scenario 5: Distributed Denial of Service Attack}
\label{subsubsec:scenario-5-ddos}
\begin{figure}
    \centering
    \includegraphics[width=0.5\textwidth]{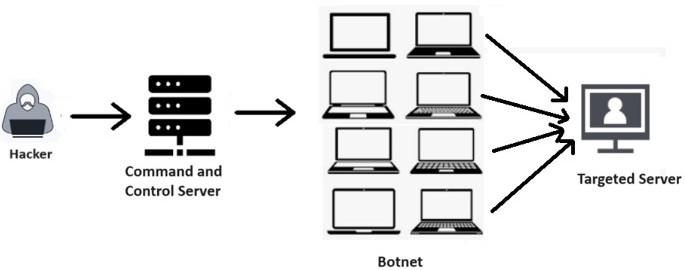}
    \caption{DDoS Attack. A botnet, controlled by a hacker, floods a target server with excessive traffic to cause service disruption.}
    \label{fig:scenario-5}
\end{figure}

In this scenario, a DDoS attack~\cite{ref:kumarasamy-ddos}, shown in Figure~\ref{fig:scenario-5}, is simulated using Kali Linux and Windows operating systems. Our attackers will be Kali Linux and Windows 7 (functioning as a bot for Kali Linux), targeting a Windows XP machine. We exploit the server message block (SMBv1) vulnerability, injecting data packets into the network. The main goal is to simulate and observe the network traffic on Windows XP, achieved through a blend of scripting, Metasploit, and external tools.

\subsubsection{Scenario 6: Hacking Group Thallium}
\label{subsubsec:scenario-6-hacking-group-thallium}
\begin{figure}
    \centering
    \includegraphics[width=0.5\textwidth]{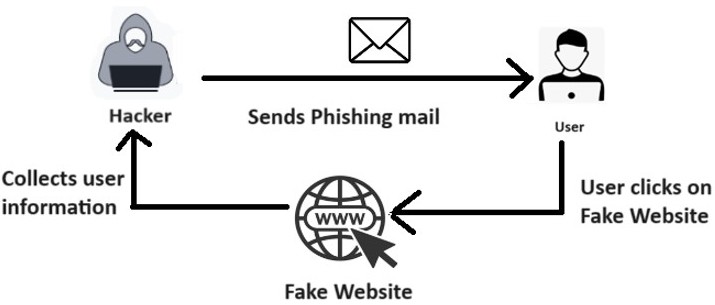}
    \caption{Phishing Attack. A user is tricked by a phishing email into providing sensitive information on a fake website.}
    \label{fig:scenario-6}
\end{figure}

This scenario, outlined in Figure~\ref{fig:scenario-6}, simulates a phishing attack using the Zphisher~\cite{ref:zphisher} tool on a Kali Linux machine, where a deceptive webpage mimicking popular websites---such as Facebook or Instagram---is created. Victims on a Windows machine interact with the phishing link, and any entered credentials are captured and displayed on the attacker's terminal. The lab also demonstrates how to make the phishing page accessible online using ngrok~\cite{ref:ngrok}, allowing real-time capture of the victim's credentials. As part of the legal reading paired with this scenario, students read the original complaint filed by Microsoft against Thallium~\cite{ref:thallium-complaint}, giving them the cybersecurity and legal background of a high-profile real-world example.

\section{Results And Analysis}
\label{sec:results-and-analysis}
A combined law and CS course composed of the scenario-based model outlined above was offered at the university in spring 2024 and spring 2025. We use the students’ performance and feedback---which consists of indirect assessment measures (interviews, surveys, and feedback) and a direct measure (lab and quiz performance)---from the course for analysis. Interviews provided qualitative insights into student experiences, while surveys covered teaching effectiveness, course content relevance, and student proficiency in cybersecurity. Student feedback mechanisms and course evaluations offered further insights to inform curriculum improvements.

\subsection{Focus Group Interview}
\label{subsec:focus-group-interview}

On March 27, 2024, focus group interviews were conducted with students from the combined law and CS cybersecurity course. Two groups, each with both law and CS students, participated in 20-minute sessions to provide feedback on the effectiveness of the course. We present the key takeaways below:

\begin{enumerate}
    \item Targeted Threats: Students valued the hands-on scenarios for enhancing their cybersecurity understanding, but CS students found them easy, while law students struggled and requested clearer instructions. Balancing difficulty and guidance would improve the experience.
    \item Holistic Knowledge: The non-technical material offered valuable multidisciplinary perspectives, though CS students desired more technical content, and law students found some readings repetitive. Refining content for both backgrounds could enhance learning.  
    \item Scenario-Based Model: While the scenario approach effectively bridged theory and practice, differing difficulty levels for CS and law students indicated a need for adjustments to equally challenge all participants.
    \item Practical Experience: Hands-on labs were well-received by CS students, and although law students faced challenges, they gained valuable exposure. Technical support improvements were suggested to address lab difficulties.  
\end{enumerate}

In addition, the students appreciated the interdisciplinary approach, which increased their confidence in cybersecurity and law, and suggested more collaboration opportunities, such as class discussions. While the course was generally successful, refining content and enhancing collaboration would further improve its effectiveness.

\subsection{Student Survey}
\label{subsec:student-survey}
As part of the course, we invited students to complete a survey at the beginning and end of the semester. The survey aimed to gauge student’s initial knowledge, expectations, and confidenfce levels (before), and to measure the impact of the course on their understanding and skills in cybersecurity (after). These responses provide insight into the effectiveness of the course in enhancing students' capabilities and confidence in tackling cybersecurity challenges. Students' responses are summarized below as they relate to the four objectives stated above.

\subsubsection{Threats Targeting SMBs}
\label{subsubsec:threats-targeting-smbs}
Prior to taking the course, students had a vague awareness of different cyber threats, with CS students having more exposure from other classes; however, both disciplines of students lacked fundamentals. By the end of the semester, students could provide concrete examples of the threat categories taught in the course. Furthermore, they reported improvement in their ability to identify vulnerabilities and apply tools to exploit them in simulated labs.

\subsubsection{Technical and Legal Understanding}
\label{subsubsec:technical-and-legal-understanding}
Before taking the course, CS students reported that they had technical confidence but lacked understanding of legal frameworks. By contrast, law students had an understanding of legal basics but were unsure how they translated directly to cybersecurity threats. At the end of the semester, CS students indicated that they could map technical requirements to legal compliance standards. Additionally, law students demonstrated that they now understood how a cybersecurity incident can trigger certain laws into effect and the importance of technical features such as logging for incident response.

\subsubsection{Scenario-Based Learning Model}
\label{subsubsec:scenario-based-learning-model}
At the start of the course, students admitted that they had little to no experience with a full incident response. Furthermore, CS and law students' experiences were fragmented on different portions of a full scenario, with CS students having some experience with technical labs and law students conducting case studies. At the end of the semester, students indicated that they had gained a step-by-step fluency in a incident response for different threats. Moreover, students reported that they had a high degree of confidence in applying legal frameworks to the different scenarios.

\subsubsection{Practical Experience and Skills}
\label{subsubsec:practical-experience-and-skills}
Students initially reported that they had no hands-on experience with incident response and only vague knowledge of the technical tools used in such cases. At the end of the course, students felt that the hands-on labs provided practical insight into cybersecurity tools and concepts. CS students enjoyed the complexity, while law students, although finding the labs challenging, benefited from the practical exposure. Finally, students reported that they felt confident in executing a clear step-by-step workflow for incident response.

\subsection{Students' Course Evalution}
\label{subsec:students-course-evaluation}
\begin{figure}
    \centering
    \includegraphics[width=0.75\textwidth]{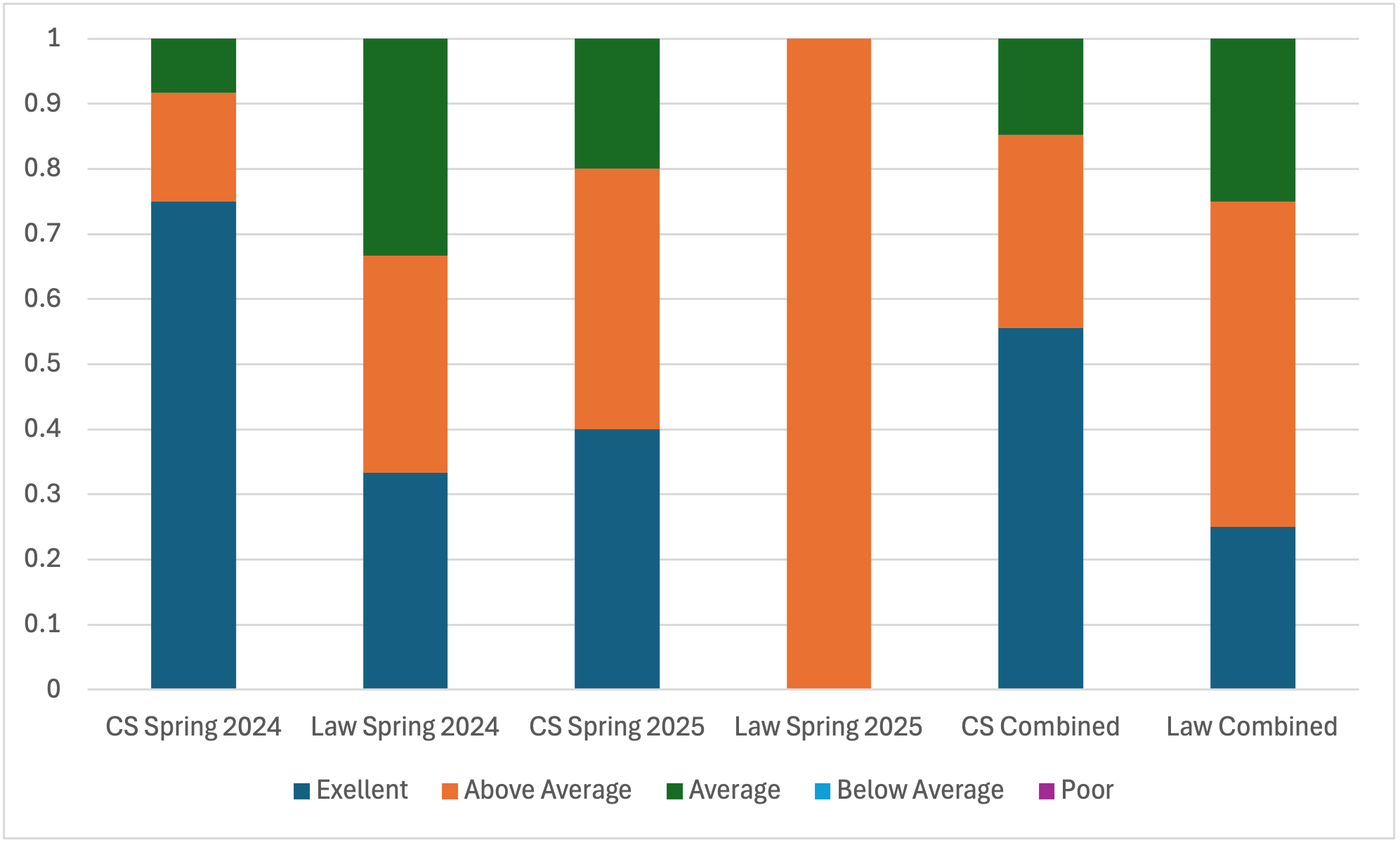}
    \caption{Students' Course Evaluation.}
    \label{fig:course-eval}
\end{figure}

As a part of the institutional mechanism to improve our teaching, students are given the opportunity to complete anonymous course evaluations. Course evaluation feedback comprises both quantitative ratings and qualitative comments from students; however, in this analysis, we focus on quantitative ratings to provide a clear numerical understanding of student satisfaction and perceived course effectiveness. Students' course ratings are shown in Figure~\ref{fig:course-eval}. We summarize the findings below.

\subsubsection{Course Evaluation from CS Students}
\label{subsubsec:course-eval-from-cs}
The quantitative course ratings from CS students indicate an overall high satisfaction with the course, with the aggregate ratings from both semesters being 4.41 out of 5. A detailed breakdown yields 55.5\% \textit{Excellent} (5), 29.6\% \textit{Above Average} (4), and 14.8\% \textit{Average} (3) as shown in Figure~\ref{fig:course-eval}. This is above both the department (4.25 in spring 2024 and 4.1 in spring 2025) and college (4.23 in spring 2024 and 4.18 in spring 2025) average course ratings. However, it must be noted that from spring 2024 to spring 2025, CS students' rating of the course decreased. This may be explained by changes in course delivery methods to better accommodate law students, indicating that there is still work to be done in finding a balance between the two disciplines. Nevertheless, when CS students' course ratings are split by semester, the average is 4.67 for spring 2024 and 4.20 for spring 2025, which is still higher than the department and college average course rating for their respective semesters.

\subsubsection{Course Evaluation from Law Students}
\label{subsubsec:course-eval-from-law}
In contrast to the CS students, law students' satisfaction with the course remained stagnant from spring 2024 to spring 2025, reinforcing the point from above that a balance must be struck to engage both types of students. The aggregate course rating of law students is 4.0 out of 5 where 25.0\% rated the course \textit{Excellent} (5), 50.0\% \textit{Above Average} (4), and 25.0\% \textit{Average} (3). The average for each semester is also 4.0. This is below the college average in both semesters (4.22 in spring 2024 and 4.16 in spring 2025), indicating more room for improvement (note that there is only one department in the College of Law, so the department average is the same as the college).

\subsection{Students' Performance}
\begin{figure}
    \centering
    \includegraphics[width=0.75\textwidth]{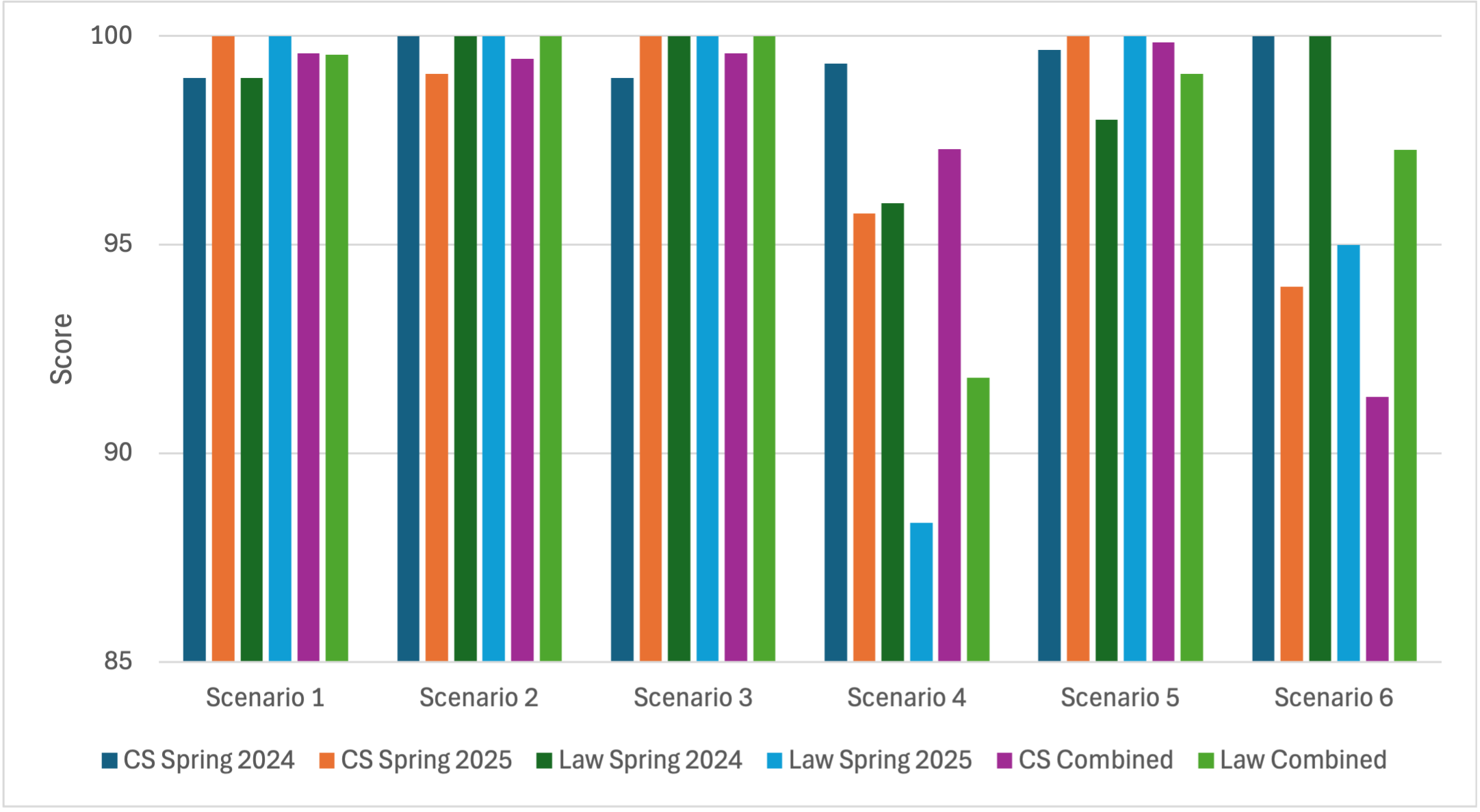}
    \caption{Average Scores of CS and Law Students on Scenario-Based Labs by Semester. Both CS and law students struggle with Scenarios 4 and 6. Note that the range on the y-axis has been condensed to 85-100 for enhanced visual clarity.}
    \label{fig:student-scores-on-scenarios}
\end{figure}

One direct measure of student learning outcomes is their performance in homework assignments (labs) and quizzes. There were six labs and six quizzes corresponding to the six cybersecurity scenarios. A detailed analysis of student performance is illustrated in the following graphs. 
Figure~\ref{fig:student-scores-on-scenarios} compares average scores between law and CS students in each lab by semester. Observe that in general, both CS and law students perform well on the lab assignments with the lowest average score being law students in spring 2025 on Scenario 4 (88.3\%). This statistic highlights the ability of the interactive scenario-based learning environment to leverage the different skillsets of law and CS students to reach a common level of understanding. That said, students had more difficulty with Scenario 4 (website fingerprinting) and Scenario 6 (phishing). In particular, for both scenarios and for both CS and law students, the performance in spring 2024 is higher than in spring 2025. It must be noted, however, that even where students experienced difficulty, the average scores for those scenarios are above 85\%. 

\begin{figure}
    \centering
    \includegraphics[width=0.75\textwidth]{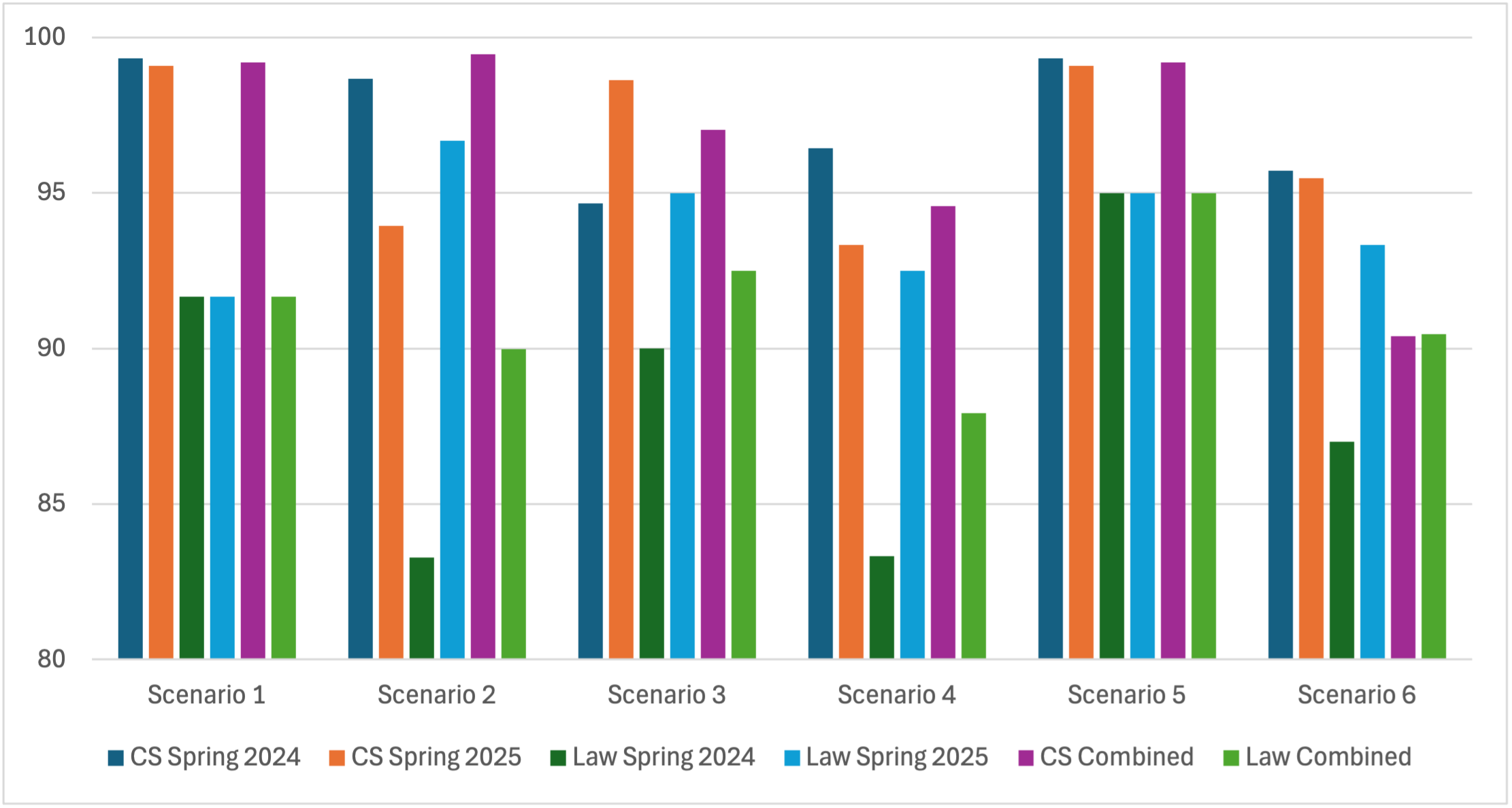}
    \caption{Average Quiz Score of CS and Law Students by Semester. Scores are variable, with CS students scoring more consistently between semesters. Both CS and law students have more difficulty with Scenarios 3 and 4.}
    \label{fig:student-scores-on-quizzes}
\end{figure}

Students' quiz scores are shown in Figure~\ref{fig:student-scores-on-quizzes}. While students still perform well on quizzes, with average scores of all students being above 80\%, the results are more variable than the labs, and law students consistently score lower than CS students. Despite this, it is important to note that law students showed a marked improvement in average quiz scores from spring 2024 to spring 2025, demonstrating an improved comprehension in the second offering of the course. This could be explained by a multitude of factors, e.g., adaptation of teaching style. We can also see that both sets of students find more difficulty with Scenarios 4 and 6, as was the case with lab scores. Such results suggest the need to revisit the content and delivery of these two scenarios, as both types of students scored lower.

To round out our grade analysis, we investigate the weighted final scores of students, shown below in Figure~\ref{fig:overall-score}. Note that scores range above 100\% due to extra credit. The results by semester are variable, with CS students scoring very consistently in spring 2024 and with more variability in spring 2025. Conversely, law students' scores are more variable in spring 2024 but more consistent in spring 2025. In both cases, the variability is explained by a few students scoring lower, skewing the mean from the median. In fact, across all semesters, the median final score is above 98\%. The overall spread of both CS and law students is similar when examining both semesters; however, the interquartile range (IQR) of CS students is far more condensed, indicating that law students have a higher degree of variability in scores. Despite differences in the IQR, the average and median final scores of law and CS students are close---the average scores are 98.5\% for CS students and 98.6\% for law students while the median scores are 95.4\% for CS students and 95.3\% for law students. Such results further exemplify the successful implementation of an interdisciplinary and hands-on scenario-based learning environment at conveying legal and cybersecurity concepts to students from different backgrounds—more explicitly, CS students learned and applied legal concepts as did law students with cybersecurity concepts. Furthermore, that law students received final grades comparable to CS students shows that while they may have experienced some difficulty with technical aspects of the course---i.e., labs and quizzes—it did not hinder their overall performance.

\begin{figure}
    \centering
    \includegraphics[width=0.75\textwidth]{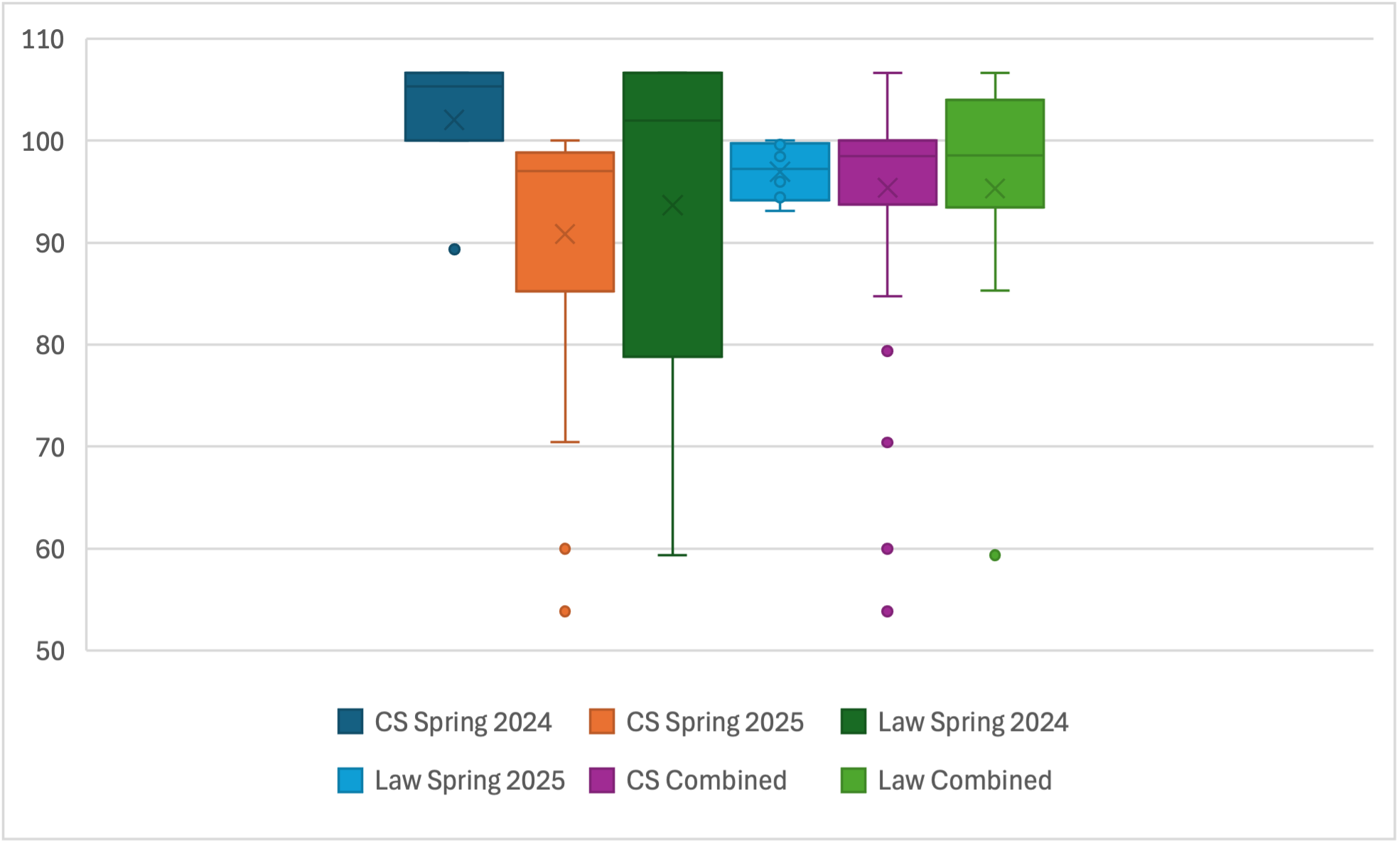}
    \caption{Distribution of Final Grades for CS and Law Students. The average and median scores for both student types are comparable. Note that a score above 100\% is possible due to extra credit.}
    \label{fig:overall-score}
\end{figure}

\section{Conclusion}
\label{sec:conclusion}
In conclusion, our research enhances cybersecurity education by developing a curriculum tailored for small and medium-sized businesses (SMBs) that integrates practical experience and legal knowledge. Feedback from students highlighted a positive reception, emphasizing the hands-on approach and real-world relevance of lab exercises, despite some challenges faced by law students. This curriculum empowers future cybersecurity professionals to safeguard SMBs and improve digital security, while the insights gained will guide future educational enhancements, fostering a more resilient cybersecurity workforce.

Our findings led to a model that serves as a sound reference for SMBs to equip themselves with and defend against the attacks discussed in this paper. With the KSAs acting as a bridge between the implementation of the best practices, SMBs can focus on the model and create evaluation and training activities for their existing workforce. SMBs can further extend the work presented in this paper by building their own competencies, and work-roles based on the TKSAs presented here and NIST guidelines. This opens up a path for SMBs to collaborate with educational institutions to build new course work based on real world scenarios. Educational institutions can design new competencies based on the TKSA presented in this paper and build a fully modular multidisciplinary course with a scenario-based learning module to help equip the upcoming cyber-workforce~\cite{ref:ghosh-assessing-competencies-using-scenario-based-learning-in-cybersecurity}.

With cybersecurity being an ever-growing and ever-changing field, KSAs can be derived for more than just the three attacks discussed in this research. In the future, our goal is to determine a list of TKSAs for cybersecurity best practices related to the Internet of Things (IoT) domain. Businesses and individuals are adopting smart technologies and many are not aware of the threats they expose themselves to. With very little literature available in the field, we aim to touch upon these topics and expand our current work to include them in the future.

\section*{Declaration of Generative AI and AI-assisted Technologies in the Manuscript Preparation Process}
During the preparation of this work the author(s) used ChatGPT in order to summarize the students' self knowledge assessment at the beginning and end of the course as they relate to the four research objectives of this work. After using this tool/service, the author(s) reviewed and edited the content as needed and take(s) full responsibility for the content of the published article.

\bibliographystyle{elsarticle-num}
\bibliography{references}

\end{document}